\documentclass[a4paper,twocolumn,showkeys,floatfix,aps,pre,reprint,groupedaddress,nofootinbib]{revtex4-1}
\usepackage{color}
\usepackage{amsmath,amssymb}
\usepackage{graphics,graphicx}
\usepackage{dcolumn,bm}
\usepackage{psfrag}
\usepackage{enumerate}

\begin{document}

\title{Social Inequality Analysis of Fiber Bundle Model Statistics and
Prediction of Materials Failure}

\author{Soumyajyoti Biswas${}^{1}$}
\email{soumyajyoti.b@srmap.edu.in}
\author{Bikas K. Chakrabarti${}^{2,3,4}$}
\email{bikask.chakrabarti@saha.ac.in}
\affiliation{
${}^1$Department of Physics, SRM University - AP, Andhra Pradesh - 522502,
India \\
${}^2$Saha Institute of Nuclear Physics, Kolkata - 700064, India \\
${}^3$S N Bose National Center for Basic Sciences, Kolkata - 700106, India
\\
${}^4$Economic Research Unit, Indian Statistical Institute, Kolkata 700026,
India
}

\date{\today}

\begin{abstract}
Inequalities are abundant in a society with a number of agents 
competing for a limited amount of resource.
Statistics of such social inequalities are usually
represented by the Lorenz function $L(p)$, 
where $p$ fraction of the population possesses
$L(p)$ fraction of the total wealth, when the 
population is arranged in the ascending order of their wealth.
Similarly, in scientometrics, such inequalities can be represented
 by a plot of the citation count 
against the
respective number of papers by a scientist, again arranged
in the ascending order of their citation counts.
  Quantitatively, these
inequalities are captured by the corresponding
inequality indices, namely the Kolkata $k$ and the
Hirsch $h$ indices, given by the
fixed points of these nonlinear (Lorenz and
citation) functions. In statistical physics of
criticality, the fixed points of  the
Renormalization Group generator functions are
studied in their self-similar limit, where its
(fractal) structure converges to a unique form (macroscopic in
size and lone). The statistical indices in the social
science, however, correspond to the fixed points
where the values of the generator function (wealth
or citation sizes) are commensurately abundant in
fractions or numbers (of persons or papers). It
has been shown already that under extreme
competitions in the markets or in the universities,
the $k$ index
approaches a universal limiting value, as the
dynamics of competition progresses. We introduce
and study these indices for the inequalities of
(pre-failure) avalanches, given by their nonlinear
size distributions in the Fiber Bundle Models
(FBM) of non-brittle  materials. We show how a prior knowledge of the
terminal and (almost) universal value of the $k$
index for a wide range of disorder parameter, can help in predicting 
an imminent catastrophic breakdown in the model. 
This observation has also been complemented by
noting a similar (but not identical) behavior of
the Hirsch index ($h$), redefined for such
avalanche statistics.
\end{abstract}


\maketitle


\section{Introduction}

The collective dynamics of failure or breaking
in any non-brittle material sample  proceeds
through the  failures of individual elements
of the material, as the external load or stress
on the sample grows. The bursts of elastic
energy released (experimentally detected as acoustic emissions)
until the complete breakdown of the material, are
widely studied (see e.g., \cite{ref7}) for the
universal nature of their statistics across length and energy scales.  These
bursts or avalanches are often also studied in
models, both analytically and numerically.
An avalanche is the sequence of failure events taking
place in the system in going from one stable state to the next,
when the external load on the system is gradually increased.
 For example, in the Fiber Bundle Model or
FBM (see e.g., \cite{ref7,ref8,ref9}), which is an ensemble
of elements having different failure thresholds collectively carrying a
global load,
an avalanche size is defined as the total number
of elements failing, immediately or due to the internal stress redistribution continued until a stable configuration is reached, after the external load
is increased on a stable configuration of the model. The avalanche size could also be
measured by the amount elastic energy released
from these failed elements. Its distribution
would then correspond more naturally to the
elastic emissions. For simplicity, however, we
consider here the avalanche size to be given
only by the number of failed elements. For
successive increases in the external load,
further avalanches of different sizes occur with
various frequencies. The probability distributions of the
avalanche sizes, across a broad class of systems,
show the common feature of having relatively larger number of smaller
events and much fewer number of large ones. Usually the
biggest avalanche is proportional to the system size and causes the
eventual macroscopic failure of the sample.

\begin{figure*}
\includegraphics[width=18cm]{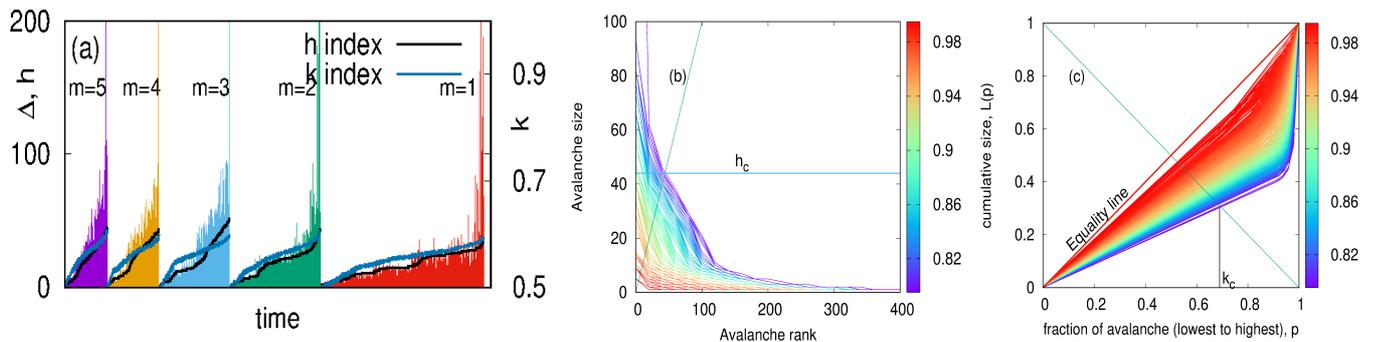}
\caption{The inequality measures for the
avalanche sizes in an FBM (with $N=50000$
fibers). (a) The time series of the avalanche
sizes ($\Delta$) are shown for different
Weibull modulus ($m$) values. The solid lines
indicate the $h$ index and the $k$ index. The
left hand side scales are for the avalanche sizes ($\Delta$) and $h$, while
the scale for $k$ index is shown on the
right hand side. While different samples continue the dynamics for different
duration, indicating different values of
the critical load $\sigma_c$, the terminal values of these two indices
vary only
weakly with $m$. (b) The rank plot of the avalanche
statistics at different stages of the failure dynamics. The colors
indicate the
fraction of surviving fibers at the time of measurement.
The 45 degree line intercept gives the $h$ index value. (c) The Lorenz
curve for
different stages of the failure dynamics. Again, the colors
indicate the fraction of surviving fiber, hence the stage of the dynamics.
The
equality line is shown, from where the Lorenz curve gets deviated
as the dynamics progresses. The terminal value ($k_c$) of the $k$ index is
indicated.}
\label{inequality_measures}
\end{figure*}

This inequality of ranks for the avalanches
is similar to what is known to exist in
societies with competing agents, for example
in the distribution of wealth among individuals,
distribution of citations of papers by
an author, and so on. There are fewer number of
rich people, like fewer papers with high
citations. There are commonly used indices
to characterize such social inequalities. These measures do not
focus on the extreme limits of the corresponding
distributions (where frequencies are either very
small or very high), but generally focus on its
typical attribute, having high value  with
commensurate value of its frequency of
occurrence.
This is in contrast to what is usually studied \cite{ref7,ref8, ref9} in
the avalanche size ($\Delta$) distribution ($D(\Delta)$)
statistics viz, the self-similar fixed
point limit of asymptotic avalanche size to
extract the limiting singularities in $D(\Delta)$.
Social inequalities  are characterized by the
fixed points  of the respective nonlinear inequality
distributions or functions. In particular, the
social  indices  like the Kolkata index $k$ (\cite{ref5},
see \cite{ref6} for a review) or the Hirsch index
$h$ (\cite{ref1}, see \cite{ref2} for a review)
correspond to the bare (or unnormalized)
nonlinear  inequality  functions, like the
complementary Lorenz function \cite{ref6} for
$k$ or the citation function  \cite{ref1} for $h$. These index
values do not directly address
the extreme, most efficient or vulnerable features
of the social or personal  behavioral statistics.
For that reason, these measures can be remarkably
stable with respect to the various parameters of the distribution
functions.

In this work, we apply the measures of social inequalities to 
the time series of avalanches of an externally stressed disordered material.
The avalanches are of different size, leading to a variation
of these indices with the progressive number of avalanches. The terminal
values of these indices correspond to the inequality measure in the system
after the catastrophic breakdown has happened i.e., there can be no further
avalanche in the system. As mentioned above, due to the fact that these
measures do not correspond to the extreme limits of the avalanche distributions,
the terminal values of these indices are very less sensitive to different
parameters of the avalanche distributions.
Therefore, these social inequality measures,
when applied to the case of a stressed disordered material showing
avalanche dynamics,
can help in characterizing the proximity of the
material from catastrophic failure point by
reflecting the emerging inequality in the avalanche statistics.

In what follows, we first define the various social inequality measures
(the Kolkata
index $k$ and the Hirsch index $h$)
and outline the way in which such measures can help in predicting imminent
catastrophic breakdowns in disordered materials.
Then we present numerical simulations of the FBM in performing the scaling
analysis
of $k$ and $h$. We then use these
analyses in quantifying the efficiency of these indices in estimating
breakdown
points. Finally we discuss the results and conclude.

\section{Method}
Here we describe the simulations of fracture of stressed disordered
materials using FBM, showing avalanche dynamics. We then describe the
methods of calculating the inequality indices ($k$ and $h$) from the
avalanche statistics. Then we outline how it is used to predict imminent
breakdown.

In the FBM, a macroscopically large number of parallel
Hooke springs or fibers are clamped between two
horizontal platforms; the upper one helps
hanging the bundle while the load hangs from
the lower one (see e.g., \cite{ref7,ref8,ref9}).
The springs or fibers are assumed to have
identical spring constant, though their
individual breaking strengths are assumed
to be different (given by a distribution). Once the load per fiber exceeds
its own breaking threshold, it fails and this extra
load is shared by the surviving fibers. If the
platforms are assumed rigid, there is no local
deformation around a failed fiber (and no stress
concentration around the `defect' created by the
failed fibers). The load is shared equally by all
the surviving fibers. We consider here this
Equal Load Sharing (ELS) scheme \cite{ref7,
ref8,ref9} for redistributing the extra load
among the surviving fibers. If this
extra load per fiber exceeds the threshold
strength of any of the surviving fiber, the
avalanche continues. The number $\Delta$ of
all the fibers failing in one go (without any
increase in  the external load) defines the
avalanche size and we study the frequency
distribution $D(\Delta)$ of avalanches as the
external load is increased until complete
failure of the bundle.

We then extract the values of  the Kolkata
index ($k$) from the fixed point \cite{ref6,ref5}
of the normalized nonlinear complementary
Lorenz function. In the context of economic inequality
of a country, the
Lorenz function (see e.g., \cite{ref6}) represents the
cumulative fraction $L(p)$ of wealth
possessed by the fraction $p$ of people
of the country, when the people
are arranged in the ascending order of their wealth. If every
person possesses equal wealth,
then the Lorenz curve (or the Lorenz function
$L(p)$) becomes the diagonal line - called the equality line - from the origin
of a unit square. But this is not what is observed. Since poor people
  have lower wealth, the Lorenz curve is nonlinear. Starting form the
origin ($L(0) = 0$), it remains below the equality
line and monotonically increases to unity at
$p = 1$ ($L(1) = 1$). Now, the Kolkata index ($k$)
corresponds to the fixed point (see e.g.,
\cite{ref3,ref5,ref6}) of the complementary
Lorenz function $\widetilde{L}(p) \equiv 1- L(p)$:
$\widetilde{L}(k) = k$. As such, it is a normalized social
inequality measure and it generalizes the
century old Pareto 80-20 law \cite{ref3}: It gives
the fraction $(1 - k)$ of people of the
country  who  collectively
possess $k$ fraction of its total wealth. This index value $k$  is
observed to approach a constant, usually higher than
0.80 (the Pareto value,  noted about a
century ago), in many social contexts, as the
collective  competitive dynamics of the society
progresses \cite{ref3}. A similar study can be
done by looking at the inequality of the citations
of the papers by an author.

For the breaking
dynamics of FBM, we numerically evaluate the
avalanche statistics $D(\Delta)$, as the internal
dynamics of (local) failures make progress.
To extract the $k$ index value at any time $t$
after the start of loading the system and before the complete failure
of the FBM
or sample,
we evaluate the Lorenz function
$L(p)$ by estimating first the fraction $p$ of
avalanches of size from 0 to $\Delta$ from the
integral

\begin{equation}
p = \int_0 ^ {\Delta} D(\delta) d\delta/
[\int_0 ^ {\infty} D(\delta) d\delta],
\label{eq1}
\end{equation}

\noindent then solving $\Delta$ as a function of $p$,
and inserting that in the expression for cumulative
avalanche size fraction

\begin{equation}
L =  \int_0 ^ {\Delta} \delta D(\delta)
d\delta/[\int_0 ^ {\infty} \delta D(\delta) d\delta].
\label{eq2}
\end{equation}


\noindent From this Lorenz function $L(p)$ (with
$L(0) = 0$ and $L(1)= 1$), we determine the $k$
index value by solving for the fixed point $\widetilde{L}(k) = k$
of the complementary Lorenz function
$\widetilde{L}(p) \equiv 1 - L(p)$.  This index
value ($k$) characterizes the avalanche
distribution $D(\Delta)$ at that time
($k$ has minimum value equal to 0.5, when the
 Lorenz curve becomes the equality line or
$L(k) = k = \widetilde{L}(k)=1 -k$, and has maximum
value equal to unity).

In the scientometrics context, the Hirsch index
$h$  corresponds to the number ($h$) of papers,
each having commensurate number ($h$ or more)
of citations at the present or running point of
the author's carrier. Typically,
if one plots the number of citations received
against the number of papers, arranged in the descending order of
citations,
the (nonlinear) plot
becomes convex towards the origin, and $h$ index
corresponds to the fixed point (intersection
point of  the 45 degree line) of this nonlinear
function. The value of the $h$ index reflects
the author's success (citation) rate in their
commensurately prolific range and not in the most
successful limit (where the author is necessarily
not prolific; highly appreciated or cited papers
are low in number!). The $h$ index helps to
distinguish among the authors, working on similar
topics, by comparing  their  success rates in
their commensurately  prolific range. When
statistically analyzed, one finds some universal
scaling behavior: $h (\sim \sqrt {N_{pap}}$ \cite{ref3}, or
$\sim \sqrt {N_{cit}}$ \cite{ref4}) where $N_{pap}$ and $N_{cit}$
denote respectively the total number of papers
written or the total citations received by the
author.

In the context of avalanche statistics,
$D(\Delta)$, we extract the failure $h$ index
and find the scaling relation for it's terminal value $h_c$ as

\begin{equation}
 h_c = C [{\sqrt N}/ log N],
\end{equation}

\noindent where the
prefactor $C$ is a function of the Weibull
modulus, characterizing the fiber strength
disorder in the bundle.

\begin{figure}
\includegraphics[width=9cm]{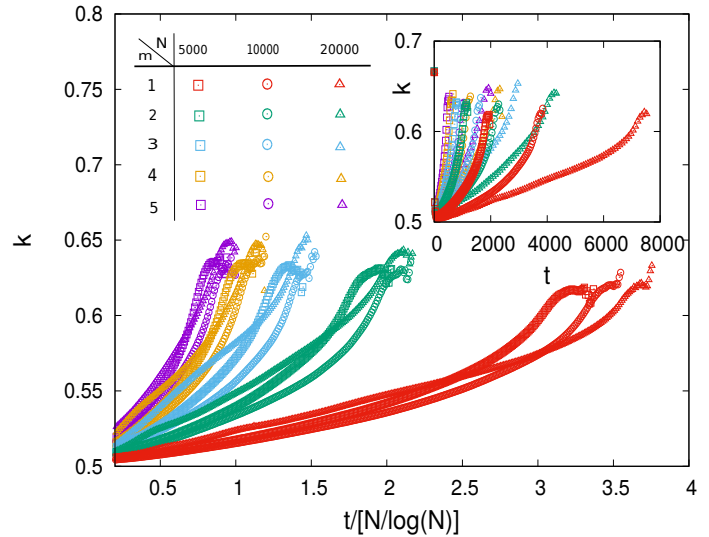}
\caption{The Kolkata index $k$ for the avalanche
distribution $D(\Delta)$ as the dynamics of failure continues in the FBM (in
the ELS scheme), where the individual fiber thresholds drawn from
Weibull distribution (1). The estimated values of the
index $k$  at different times $t$ (scaled by $N/log N$) are plotted until
complete failure of the bundle (with  disorder characterized  by different
 Weibull  moduli ($m$) indicated using different colors). The terminal
value of the
$k$-index, prior to complete failure bundle, seems reach a threshold ($0.62
\pm 0.03$) and this terminal value is weakly dependent on $m$.  The inset
shows the
variations of $k$ index with unscaled time.}
\label{kindex_weibull}
\end{figure}

\begin{figure}
\includegraphics[width=9cm]{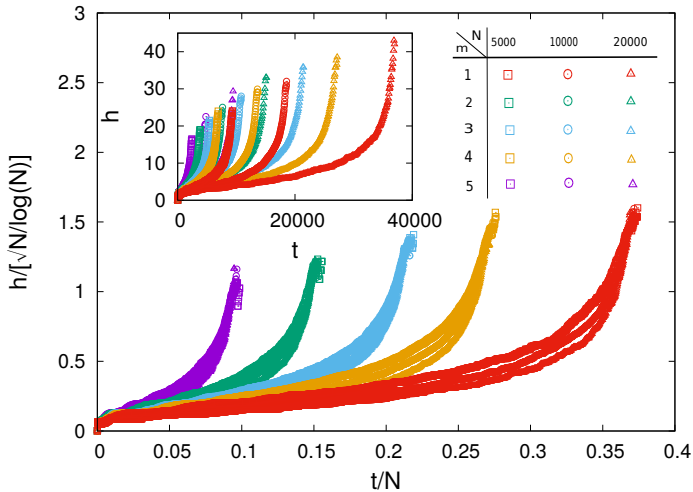}
\caption{System size ($N$) scaling of the failure index ($h$)
as the dynamics of failure occurs in the FBM (in the ELS scheme)
where the individual fiber thresholds drawn from
Weibull distribution (1). The estimated values of the
failure  $h$ index (from numerical evaluation of $D(\Delta)$)
at different times $t$ are plotted until complete failure of the
bundle (with  disorder characterized  by different  Weibull
moduli ($m$) indicated using different colors). The $h$-index values prior
to complete failure bundle scales as $\sqrt{N}/(log N)$,
and its  limiting value (just before breaking) varies only with
$m$. Our study demonstrates that  prior knowledge  of this
limiting constant would help predicting the failure
time (scaled by the bundle size). The inset shows the
variations of unscaled $h$ index with unscaled time. }
\label{hindex_weibull}
\end{figure}

The two indices defined above are monotonically
increasing functions of time $t$
in an avalanching system. The values reach some terminal or critical limit
 ($k = k_c$ and $h = h_c$) before the catastrophic breakdown of the
material. As mentioned before,  these terminal values are indicative of
the emerging inequalities in the dynamics and
not based on the extreme events only. These values are, as we shall
demonstrate in the following section,  remarkably stable with respect to
various threshold distributions of the FBM. Therefore, monitoring the
growth of $k$ and $h$ values and stopping the loading process before the
average terminal values $k_c$ and $h_c$ in the FBM, can help in loading a
system and minimize the risk of overloading (hence breakdown).

\section{Numerical Studies for $h$ and $k$ indices in FBM}

We consider here a FBM system consisting of $N$ fibers
($5,000 \le N \le 100,000)$ having identical spring
constant,  but having different failure
strengths $\sigma_f$ given by the cumulative Weibull distribution

\begin{equation}
 F(\sigma_f) = 1  - exp [-(\sigma_f)^m].
\end{equation}

\noindent Here $m$ denotes the Weibull modulus and we
choose the range $1 \le m \le 5$ for this study.
The variation in the Weibull modulus for the simulations takes care
of the fact that in real samples the disorder strengths can be very different.
A larger value of $m$ makes the failure more abrupt i.e., a shorter time series of avalanche -- in 
a limiting case leading to a brittle failure. A smaller $m$ would eventually lead to 
individual, single failures. In between these two limits, there exists an avalanching quasi-brittle 
regime, where a temporal correlation exists in the breaking statistics which is commonly seen in experiments. 
In choosing the range for the Weibull parameter, therefore, we
made sure that the dynamics of the model is in the quasi-brittle region,
showing scale free avalanche size distribution. For $m<1$, the threshold
distribution
becomes very wide and the failure progresses through individual fiber breaking without any temporal 
correlation or large avalanche. For large values of
$m$, the
system becomes brittle i.e., the first avalanche breaks the entire system.

\begin{figure}
\includegraphics[width=9cm]{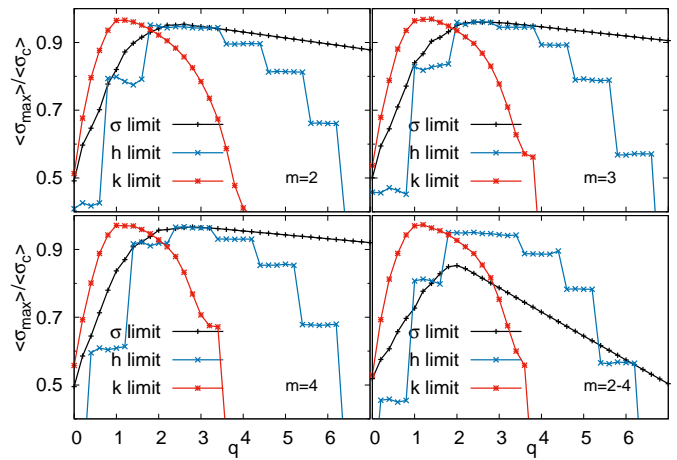}
\caption{The average safety limit of load $W_{max}^{est}$
${(\equiv N <\sigma_{max}>)}$ on an $N$-fiber system
(for 3 types of sample sets with distinct values of Weibull modulus $m$, and
one mixed type of sample set with values of $m$ in the range 2-4),
estimated using the knowledge about the terminal values $k_c$ or $h_c$ of
the Kolkata index $k$ or the Hirsch index $h$. Estimate of the
maximum load capacity on the bundle $W_{max}^{est}$ is made
by checking either the $k$ index value reaching $k_c-q\chi_k$ or
or the $h$ index value reaching $h_c-q\chi_h$.
In all these cases, the safety limit of loading is the highest for the
estimate using the knowledge  of $k$ index terminal value.
For comparison, we also make these estimates by monitoring the load per
fiber and stopping before the load
$\sigma_c-q\chi_{\sigma}$. The estimate using $k$ is higher than this also.}
\label{max_load}
\end{figure}

The external force (stress) on the FBM increases
until a fiber fails and does not increase further
until the successive fibers fail due to stress
readjustments. As mentioned before, the number
$\Delta$ of such failed fibers  in one go
(before the stress is increased further)
defines  the avalanche size. The external
load is then increased further until the weakest surviving 
fiber(s) fail and causes a further avalanche.
Effectively this means that the external load
on the bundle increases very slowly, since the
load readjustments following any fiber failure
are very fast. The process then
continues until the entire bundle fails.
We study the (frequency) distribution $D(\Delta)$
of the avalanche sizes  $\Delta$ until the time $t$
(where $t$ = 0 corresponds to the time of
putting load on the bundle)
and continue up to complete failure
of the bundle. At each intermediate
point of time $t$, we estimate the $k$ index
value at $t$ by evaluating first the Lorenz
function  $L(p)$  following the equations (1)
and (2) and then finding the fixed point $k$
of $\widetilde{L}(k) \equiv 1 - L(k) = k$.  We also
determine the $h$ index values as given
by the size ($\Delta =h$) of the avalanche which
matches its frequency ($D(h) = h$) of occurrence. We
average typically over 100 to 10,000 disorder
configurations.

Fig. 1 shows how the values of different dynamical quantities
(including those for indices $k$ and $h$)  change as
the dynamics of breaking progresses in some representative FBMs.
The estimated values of the  index $k$  at different
times $t$ (scaled by $N/log N$) before the bundle fails,
are plotted in Fig. \ref{kindex_weibull}, until complete failure of  the
bundle (where the  disorder of the fiber of the bundle,
characterized  by different  Weibull moduli ($m$),
is  indicated using different colors). The observed
terminal values of the $k$-index ($k_c= 0.63 \pm 0.02$),
prior to  the complete failure of the bundle seem to be
weakly dependent on $m$).  Similar universality, but
at a higher terminal value, was seen in the cases of
citation index $k$  of authors in the limit of
extreme competitiveness \cite{ref3}.
Our study demonstrates that  prior
knowledge  of this limiting value of the Kolkata
index $k$ for the growing avalanche statistics
$D(\Delta)$ would thus help predicting the
imminent failure point or time (when scaled by $N/log N$).

In Fig. \ref{hindex_weibull} we show that the failure $h$ index of the FBM
scales as in Eq. (3) with a $log N$ correction
factor to the scaling behavior $h \sim \sqrt N$,
also seen in the context of journal citations \cite{ref3, ref4} .

\begin{figure}
\includegraphics[width=9cm]{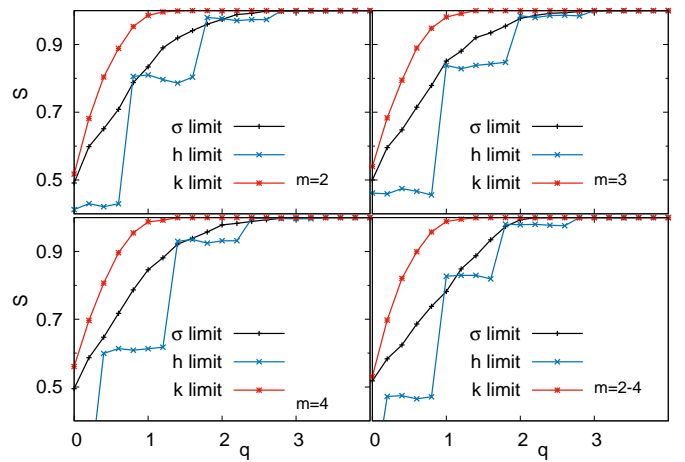}
\caption{The average fraction of samples surviving due to the loading
process shown
in Fig. \ref{max_load}.
Towards the low values of $q$, almost half the samples are breaking due to the
applied load. For
high values of $q$, all the samples survive.}
\label{surv_frac}
\end{figure}

In demonstrating how the social index, for example, $k$ ($h$) could be
useful in  predicting the imminent failure, we load a system and
continuously monitor its index value, until the value is a multiple ($q$)
of the standard deviation $\chi_k$ ($\chi_h$) away from the
average critical (terminal) value  $k_c$ of $k$ (or $h_c$ of $h$). The
system of course  can break before that, due to sample-to-sample
fluctuations of these terminal values, and in those cases we then estimated
load carrying capacity ($W_{max}^{est} \equiv
N\langle\sigma_{max}\rangle$) to be zero
in its statistics. As can be seen in Fig. \ref{max_load}, for very low
values of $q$,
the loading is not stopped until the average terminal value is reached,
causing breakage of almost half of the samples (see Fig. \ref{surv_frac}),
making
the estimated
safety limit of $W_{max}^{est}$ much lower.  For very high values of $q$,
the loading is stopped too soon, again making $W_{max}^{est}$ too low. In
an intermediate range of $q$,  the value of $W_{max}^{ets}$ reaches a
peak. The peak value is the highest when the  estimate of $W_{max}$ is
made using the knowledge of the $k$ index terminal value, making it the
most effective  monitoring parameter for safe-loading.  The other
parameters discussed  here (or any other method we know of) give lower
estimate of the loading capacity $W_{max}$ for samples characterized by
Weibull distributions (at least in the range of the modulus $m$
considered here). We also checked this method for uniform distribution of
the failure threshold in the range $(0.5-r,0.5+r)$. When the system is
close to the brittle limit $r=1/6$, the predictability using $k$ index is
less effective. However, for higher values of $r$ (see Fig. 6), $k$ index
based loading works best, as before.


\begin{figure}
\includegraphics[width=9cm]{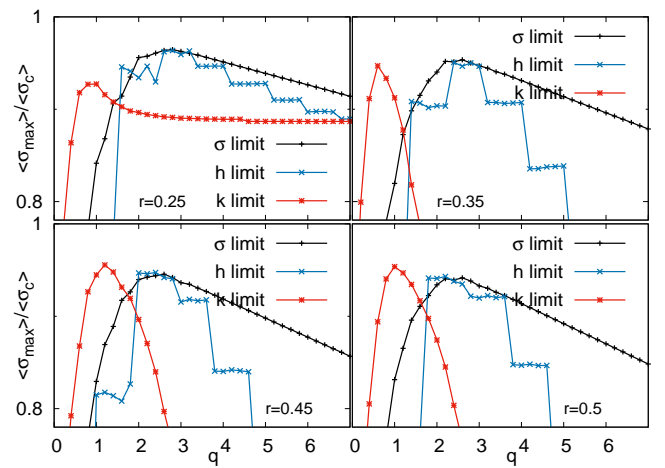}
\caption{The average safety limit of load $W_{max}^{est}$
${(\equiv N <\sigma_{max}>)}$ on an $N$-fiber system
with threshold distribution taken as uniform between $(0.5-r,0.5+r)$. In
most cases, loading by monitoring $k$ give maximum load (except near the
brittle
limit).}
\label{max_uni_load}
\end{figure}

\begin{figure}
\includegraphics[width=9cm]{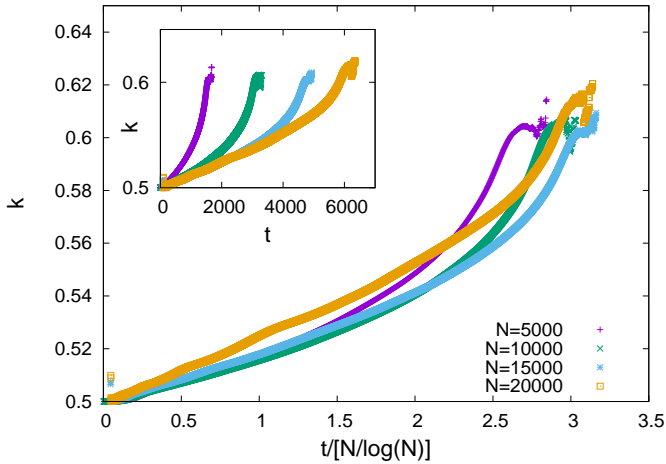}
\caption{The system size scaling of the $k$ index for threshold
distribution uniform
in $(0,1)$. The terminal value of $k$ becomes $0.62\pm 0.03$.}
\label{k_index_uni}
\end{figure}

\begin{figure}
\includegraphics[width=9cm]{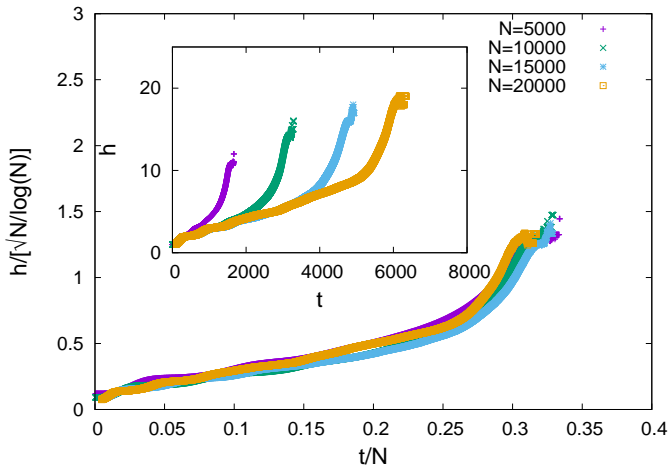}
\caption{The system size scaling of the $h$ index for threshold
distribution uniform
in $(0,1)$.}
\label{h_index_uni}
\end{figure}

\begin{figure}
\includegraphics[width=9cm]{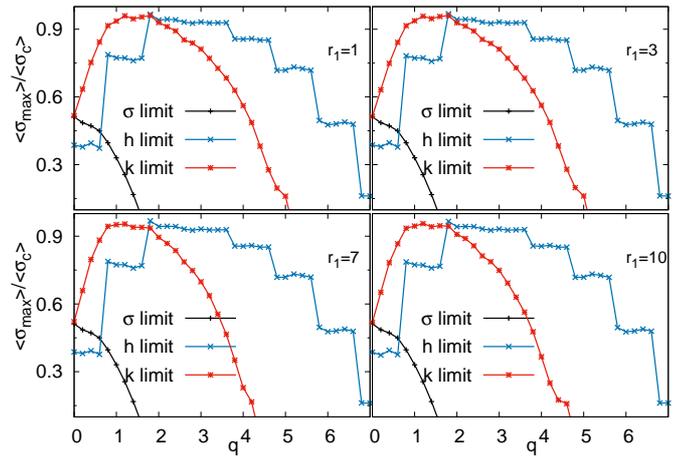}
\caption{The maximum possible loading, scaled by the corresponding
critical loads,
when the
threshold distributions are uniform in the range $(0,r_0)$, but the value
of $r_0$
is distributed
uniformly in the range $(0,r_1)$.}
\label{unidist_2}
\end{figure}

In order to verify the universality of the system size scaling of $k$ and $h$
indices in FBM
and also to further justify the effective monitoring process of
safe-loading, we
study the
dynamics of the model when the threshold distribution is uniform between
$(0,1)$. In
Figs. \ref{k_index_uni} and  \ref{h_index_uni},
the system size scaling of $k$ and $h$ indices are shown and they seem to
obey the
same scaling
as was noted before for the Weibull distribution (Figs.
\ref{kindex_weibull} and
\ref{hindex_weibull}). To demonstrate the safe-loading,
we choose sample-sets where the threshold distributions are uniform in
$(0,r_0)$,
but the value of $r_0$
is again chosen from a uniform distribution within $(0,r_1)$. In that
case, the
critical load $\sigma_c$
will vary strongly with the upper limit of the distribution, but the
inequality
indices will not.
Fig. \ref{unidist_2} shows the corresponding maximum loading for different
values of
$r_1$. In all cases, monitoring
$k$ and $h$ indices perform much better than monitoring $\sigma$. Indeed,
as is seen
in the system size
scaling of the $h$ index, it will vary depending upon the system size (Fig.
\ref{h_index_uni}), while no such systematic
variation exists for the $k$ index (Fig. \ref{k_index_uni}). It is
therefore, the
most useful monitoring parameter for
safe loading studied here, as its terminal (critical) value does not have any
systematic dependence on
the size of the system and a rather weak dependence on the parameters of the
threshold distributions, as long as
the failure dynamics is not too close to brittle failure.

\section{Summary \& Discussions}

The breaking dynamics of disordered materials under
slow external loading proceeds through intermittent
bursts of avalanches of a wide range of sizes. There
is no characteristic size i.e., the size distribution
of these avalanches are scale-free and typically follows
a decaying power-law behavior in the asymptotic limit of
large avalanche sizes. What intrigued researchers over
the last three decades, is the emerging universality
of the exponent value of the avalanche size distribution
from the tectonic scale of the earthquakes (Gutenberg-Richter
law) to the laboratory scale of quasi-brittle materials.
Given the striking regularities in such statistics,
both in experiments and in numerical model simulations,
various features of avalanche size distributions have been
routinely used in aiming to predict imminent catastrophic
avalanches (see e.g., \cite{scholz,kun,epl,zaiser,lasse,scirep}).

We have studied here the failure  dynamics of
non-brittle materials using the
Fiber Bundle Model, having  fiber strengths
characterized by Weibull distribution (4) and also
the case of uniformly distributed fiber threshold.
Specifically, we have studied
here numerically the avalanche distribution
$D(\Delta)$ of the avalanches of size $\Delta$ as
the dynamics of breaking proceeds. The different values of the
Weibull modulus correspond to the differences
in individual samples, as is also seen in 
citation counts of individual authors or 
wealth distributions in economies of different 
countries.

As mentioned already, the critical behavior
(characterized by
the critical exponents) very near the critical or
breaking point of the bundle (where the avalanche
size $\Delta$ reaches its asymptotic limit
of O($N$)) is very well studied
(see  e.g., \cite{ref7,ref8,ref9}). Indeed,
the critical behavior of the avalanche distribution
$D(\Delta) \sim \Delta^{-\gamma}$, and the
universality class given by  the exponent
$\gamma$ in this FBM (with equal load sharing),
have also been studied in other widely
different contexts (see e.g., \cite{ref10}).
Although the detailed knowledge of the critical
behavior for such breaking in the FBM are
extremely useful in comprehending the universality
class of breaking phenomena and their statistics
in different contexts, because of the  tiny
extent  of the critical
region (before the complete failure point, where
such critical behavior can be observed), they do
not help much in the attempts to predict the
breaking point or time  (see however \cite{ref9,ref11}
for overloaded FBM cases).

We studied here numerically (for $5,000 \le N
\le 100,000$) the avalanche distribution
$D(\Delta)$ at different times $t$ during the
failure dynamics for the bundle, under slowly
but uniformly increasing load on the bundle,
until complete failure. The failure index $h$ is
given by the avalanche size $h$ of  $\Delta$
which is equal to its frequency of occurrence
($h = D(h))$ at any time $t$ during the process
of breaking  of the bundle until its complete failure.
For extracting the  values of the Kolkata index $k$
(at different times $t$ of the dynamics) we
evaluated first  the Lorenz functions $L(p)$
(giving the cumulative fraction of avalanche sizes of
$p$ fraction of avalanches when arranged from
the smallest to the largest
or $N$ order avalanches), using Eqs. (\ref{eq1}) and (\ref{eq2}).
We then look for  the fixed
point solution of the equation $1 - L(k) = k$ of the
complementary Lorenz function $\widetilde L \equiv 1 - L$
given by the distribution $D(\Delta)$ at that time $t$.
Fig. \ref{inequality_measures}  how the values of
different dynamical quantities, in particular
the indices $h$  and $k$ change as the  dynamics
of breaking progresses in some representative
FBMs. These estimated values of $k$ at different
times $t$ of the bundle breaking process, before
the complete failure of the bundle, are plotted
in Figs. \ref{kindex_weibull} and \ref{k_index_uni}. The observed
terminal value $k_c$ ($= 0.62 \pm 0.03$) of the
$k$-index, prior to complete failure of all the
studied bundles, seems to be practically
independent of the nature of
disorder in the bundle (e,g., Weibull modulus $m$,
or the uniformity of the distribution of fiber strengths
in the bundle, or the bundle size $N$) though the breaking time
depends strongly on $m$ and  fiber number $N$ in
the bundle). It
may be noted in this connection that this average
value of $k_c$ ($\simeq 0.62$) is approximately
inverse of the Golden ratio which is the precise
value of $k$-index when the Lorenz function $L(p)$
becomes quadratic in $p$ (see e.g., \cite{ref6}). The fixed point value $k$ of
$\widetilde L(k) =k = 1 - k^2$ is then given by $k = k_c
= (\sqrt{5} - 1)/2 \simeq 0.618$.  As discussed
in the previous section, the advantage of
monitoring the $k$-index value for estimating the
maximum load on the bundle (clearly  demonstrated in
Figs. \ref{max_load}, \ref{max_uni_load} and \ref{unidist_2} are very
encouraging. We also found (see Fig. \ref{hindex_weibull}) that
the scaled prefactor ($C$) of $h$, given by
$h/[\sqrt N/log N]$, approaches some fixed limiting
values dependent on the Weibull  modulus $m$, at
the bundle failure point or time (scaled with $N$).

As we mentioned in the Introduction, in social
sciences the important indices try to capture the
structure of the inequality distributions (e.g., wealth, citations)
typically in its fixed point region, where the
distribution frequency is neither very weak (as in
the super rich and highly cited limit) nor
very prolific (as in the poor and scarcely cited limit).
As we showed here in Figs. \ref{kindex_weibull} and \ref{k_index_uni}, the
(almost)
universal terminal value of the
Kolkata index $k_c$ ($= 0.62 \pm 0.03$) for  the
statistics $D(\Delta)$ of avalanches  can help
in an unambiguous  way in predicting the complete
failure point or time of the FBM. The
scaling prefactor ($C$) of the failure $h$ index
can also help locating the macroscopic failure
point of the FBM (see Figs. \ref{hindex_weibull} and \ref{h_index_uni}),
provided the precise knowledge of $m$ and $N$
are available. This aspect of the terminal value
$h_c$ of $h$-index causes its use to be considerably
limited.

In Sociophysics (see e,g., \cite{ref12}) or econophysics (see
e.g., \cite{ref13}) it is usually argued that models and
techniques of statistical physics can lead to major
success in comprehending the social and economic
phenomena. Our study here may be the first one to show that
various statistical indices of social sciences
can in turn lead to some useful predictive power
for the dynamics of failures in materials: The
Hirsch and Kolkata indices, given by the fixed points
of the avalanche size distributions (much away from
their self-similarity induced critical breaking
point) seem to offer  some unique and  potentially
useful techniques for predicting the failure of
materials. We demonstrate the success of two such indices for the failure
statistics of materials in almost one hundred year
old Fiber Bundle Model, extensively studied both computationally and analytically in
some limiting cases (see e.g,, \cite{ref7,ref8,ref9}); and
experimentally (see e.g., \cite{lomov}).
Needless to mention
that analyzing the experimental data for the
time series of ultrasonic emissions before complete
failure in materials and similar studies for some of the
established theoretical models of self-organized
critical dynamics in FBM (see e.g.,  \cite{ref14}) and  in
earthquakes (see e.g., \cite{ref15}) will be extremely important.

\bigskip

Acknowledgment: We are thankful to Srutarshi Pradhan for
important comments and suggestions. BKC is grateful to the Indian
National Science Academy for their Senior
Scientist Research Grant.

\end{document}